# Tell me who its founders are and I'll tell you what your online community looks like: Online community founders' personality and community attributes

Yaniv Dover[a,b], Shaul Oreg[a]

[a] The Hebrew University Business School, Jerusalem, 9190501, Israel

[b] The Federmann Center for the study of Rationality, The Hebrew University of Jerusalem, Jerusalem, Israel

**Abstract**

Online communities are an increasingly important stakeholder for firms, and despite the growing body of research on them, much remains to be learned about them and about the factors that determine their attributes and sustainability. Whereas most of the literature focuses on predictors such as community activity, network structure, and platform interface, there is little research about behavioral and psychological aspects of community members and leaders. In the present study we focus on the personality traits of community founders as predictors of community attributes and sustainability. We develop a tool to estimate community members' Big Five personality traits from their social media text and use it to estimate the traits of 35,164 founders in 8,625 Reddit communities. We find support for most of our predictions about the relationships between founder traits and community sustainability and attributes, including the level of engagement within the community, aspects of its social network structure, and whether the founders themselves remain active in it.

Online communities have emerged as critical assets for marketers, serving not just as spaces for customer interaction, but as powerful engines of value creation, brand loyalty, and innovation. These communities amplify customer engagement, co-create meaning and products (Muniz Jr and O'guinn 2001; Schau, Muñiz Jr, and Arnould 2009), and boost sales through social influence and emotional bonds (Algesheimer, Dholakia, and Herrmann 2005; Bagozzi and Dholakia 2006; Manchanda, Packard, and Pattabhiramaiah 2015). Consumers immersed in these environments often form strong communal identities and act as brand advocates (Schau et al. 2009), and the online brand communities in turn also foster trust and facilitate organic word-of-mouth marketing (Godes and Mayzlin 2004; Kozinets et al. 2010). These dynamics underscore their strategic value and help explain the interest of both practitioners and researchers in understanding online communities.

To deliver value to marketers and consumers, however, online communities need to survive their initial founding, sustaining meaningful activity levels over time. Moreover, their long-term success depends on multiple factors, including their initial design, cultural coherence, and their ability to sustain enduring social structures (Brogi 2014; Fournier and Lee 2009; McAlexander, Schouten, and Koenig 2002). As such, much of the research of online communities focuses on understanding and predicting community viability (Zhu, Sun, and Chang 2016) as well as other meaningful attributes that reflect their mode of functioning (Fournier and Lee 2009). In line with this body of research, our focus in the present study is on the prediction of community attributes and success (e.g., survival over time). To date, most of the evidence into the determinants of community attributes comes from small-scale survey-based research, a few qualitative studies (e.g., Barrett, Oborn, and Orlikowski 2016; Cunha et al. 2019) and a collection of observational studies about the relationship between user activity and communities' attributes (e.g., Arazy et al. 2011; Arazy et al. 2015; Bateman, Gray, and Butler 2011; Butler et al. 2014; Lazar and Preece 2002; Lin 2008). According to this research, predictors of online community success can be generally classified into three categories: (1) measures of community activity, e.g., the level of activity soon after a community has been formed, or the impact of members' active roles (Arazy et al. 2016; Bateman et al. 2011; Cunha et al. 2019), (2) measures of the community's social network, such as social cohesion, diversity of connections (Hinds and Lee 2008; Lin 2008; Robert Jr and Romero 2017) and network centralization (Antonacci et al. 2017), and (3) measures of the effectiveness of the digital interface through which the community operates (Lu, Phang, and Yu 2011). Although these factors significantly improve our understanding of communities' success, they are far from being exhaustive, and scholars are continuously searching for new perspectives and additional predictors of community attributes and outcomes. Integrating organizational research linking founder and leader attributes to attributes of their organizations (e.g., Oreg and Berson 2018; Schneider et al. 1998) and research linking individuals' personality traits with their online behavior (Adamopoulos, Ghose, and Todri 2018; Chang, Hsieh, and Lin 2013), we propose that online community attributes can be predicted from their founders' personality traits.

In the context of online communities, the term "founder" refers to those individuals who are the first to establish and become active members of the community (aka community "pioneers"; e.g., Foote et al. 2017; Kraut and Fiore, 2017). This is a somewhat narrower meaning than that used in other organizational contexts, given that community founders' hold over their communities is typically weaker than that of organization founders over their organizations, and community founders' means of influence are less direct than in other contexts. This makes their role in shaping their communities particularly intriguing. Our focus in the present study is on assessing online community founders' personality traits and using these assessments to predict community attributes.

Overall, little is known about the role of online community founders. A few studies focused on those individuals listed as the creators of the community and tested relationships between these individuals' activity-related attributes and attributes of the community. For example, community creators' activity early on in the community's life, their activity in other communities, their motivations, and the number of their connections within the community, were all associated with the community's membership growth and its probability of remaining active months following its creation (Foote and Contractor 2018; Foote, Gergle, and Shaw 2017; Kraut and Fiore 2014). Inspired by research of organizations and their top leaders (for a review see Oreg and Berson 2018), we propose that founder personality will be another important determinant of community success.

In other organizational contexts, founders' and leaders' personality has been said to have a key role in determining organizational characteristics and outcomes (e.g., Hambrick 2007; Schneider et al. 1998). This is, in fact, a key tenet of Upper Echelons Theory (Hambrick 2007; Hambrick and Mason 1984), according to which top leaders' personality determines their behavior, which in turn sets the tone for the entire organization and ultimately determine its attributes. Through such processes, leaders in general and founders in particular are said to mold organizations in their own image (Hambrick 2007; Schneider 1987).

Although not in online communities, evidence for the relationship between leaders' personality and organizational attributes has been demonstrated in a number of studies (e.g., Berson, Oreg, and Dvir 2008; Gamache et al. 2015; Gupta, Briscoe, and Hambrick 2018; Nadkarni and Herrmann 2010). For example, organizations headed by CEOs characterized as high on security values tended to be stable and efficient, whereas those headed by CEOs high on benevolent values tended to have a more supportive environment (Berson et al. 2008). Similarly, organizations of leaders who value openness to change were more likely to be have organizational climates oriented toward innovation, and organizations of leaders who emphasize achievement values were more likely to have climates oriented toward performance goals (Berson and Oreg 2016). Similar effects have been found for organizations' top management team, beyond the effects of a single leader's personality (e.g., Colbert, Barrick, and Bradley 2014).

Recognizing the importance of leader personality, some research of online communities has been devoted to characterizing community leaders (i.e., community moderators, holding the authority to accept or remove members, and to edit content, Gazit 2021). Overall, these leaders have been characterized as extraverted, open to experiences, and emotionally stable (Gazit 2021). Despite this overall, tentative, characterization of community leaders, substantial differences exist in community leaders' personalities, and these differences can be very meaningful for predicting community attributes. Accordingly, we focus in the present study on the relationships between the personality traits of community founders and the long-term properties of their communities. Specifically, we test the relationships between founders' Big Five personality traits (Digman 1990) and attributes of their communities, such as their structure and sustainability (i.e., the likelihood that they will remain active over time).

Although findings on founder personality in other organizational contexts provide a good starting point for thinking about the impact of community founders, many of the processes through which founders are said to influence typical organizations are less relevant for explaining how online community founders influence their communities. Specifically, relative to leaders of traditional organizations, community leaders have substantially less power to reward and punish community members and less authority and legitimacy to do so. As such, relationships between online community founder personality and community attributes are much less straightforward than these relationships in traditional organizations (although even in these more traditional organizations, establishing such relationships empirically has not been trivial, see Oreg and Berson 2018). As we explain below, however, community founders can still influence their communities by functioning as role models, through processes of socialization. We describe such processes below to establish our predictions about the relationships between founder personality traits and community attributes.

Despite its potential to explain online community attributes, research has yet to consider founders' personality. One possible reason for this is the difficulty in executing such research. To study these relationships, a large number of communities needs to be sampled, their leaders' personality needs to be assessed, and the communities then need to be tracked over time to evaluate their long-term attributes. Neither of these steps are easy to execute. It is particularly difficult to obtain direct measurements, of sufficiently large samples, of community leaders' personality traits. We therefore developed a text mining approach for the assessment of online community members' personality traits, which can then be used to assess the personality traits of a large number of founders of a large number of communities. We used this approach within Reddit, one of the world's largest online community platforms, comprising 52 million daily active users in more than 100,000 active communities (Reddit). Each community in Reddit (called a subreddit) constitutes a virtual space shared by users who join it, and is dedicated to specific interests.

Building on extant text-based assessments of individuals' personality (e.g., Xia Liu, Li, and Xu 2021; Yarkoni 2010), we developed a text mining algorithm for assessing community members' personality traits by sampling 1,025 Reddit users, from whom we obtained their self-reported personality traits and their pre-existing texts on the platform. We then used this sample to create an algorithm for assessing Reddit users' personality using their public text. Using this algorithm, we estimated the Big Five traits of founders of 8,625 online Reddit communities. Importantly, the text we used to estimate founders' personality traits was produced by the founders *before* they joined the online communities we studied. We defined users as founders if they were among the first ten users to join and remain active in the community during the first month since its establishment. Finally, we examined the relationships between founders' estimated personality traits and properties of these founders' online communities, one year following these communities' inception.

We found that the likelihood that an online community will remain active a year following its inception is positively associated with founders' conscientiousness and agreeableness, and negatively associated with founders' extraversion. Among those communities that remained active after a year, founders' neuroticism was negatively associated with engagement in the community whereas agreeableness and conscientiousness were positively associated with it. Finally, founders' extraversion was positively associated with ego network size and agreeableness was negatively associated with it. In other words, members of communities with extraverted and less agreeable founders tended to interact with a larger number of peers relative to members of communities with introverted and agreeable founders. Overall, our findings demonstrate that founders' personality traits predict online community attributes, highlighting founders' important role in the growth and sustainability of these communities.

**Research Background and Theory Development**

As noted above, our focus in this study is on the relationships between online community founders' personality traits and attributes of their communities. In personality we are referring to the typical manner in which an individual thinks, feels and behaves (Cervone and Pervin 2015). Within the domain of personality, traits are perhaps the most researched aspect and are often used for predicting people's attitudes and behavior (Barrick and Mount 1991; DeNeve and Cooper 1998). Numerous such predictions have been substantiated using the Five Factor Model of personality (Digman 1990), in which personality traits are classified into five broad personality dimensions: Neuroticism (the tendency toward emotional instability, being prone to anxiety and depression), Extraversion (the tendency to be outgoing, energetic and communicative), Openness to experience (exhibiting curiosity, originality and open mindedness), Agreeableness (oriented toward cooperation, friendly and polite) and Conscientiousness (being dependable, hardworking and methodical). In numerous studies, the five factors (also known as the Big Five) have been linked with people's behaviors and choices (e.g., Hirsh, DeYoung, and Peterson 2009; Larson, Rottinghaus, and Borgen 2002).

Alongside research about the effects of personality on an individual's own behavior, a growing body of evidence demonstrates the effects of founders' and top leaders' personality on their constituents and the organizations they lead (for a review see Oreg and Berson 2018). Indeed, founders and senior leaders have been said to shape organizations in their image (Hambrick 2007). Their personality determines their decisions—such as whom to recruit into the organization, and what types of behaviors to reward—which in turn determines the organization's attributes. Accordingly, top leader personality has been shown to reflect in the organization's climate and culture (e.g., Berson et al. 2008). In other research, top leader Big Five traits were linked with companies' financial outcomes (e.g., Colbert et al. 2014; Wang et al. 2016). For example, CEO emotional stability, openness to experience and conscientiousness were linked with improved organizational performance (e.g., Colbert et al., 2014). In another study, the personality of new venture founders was linked with the emergence of conflicts among the top management team and the new venture's performance (de Jong, Song, and Song 2013). Among the findings, founders' neuroticism was positively, and extraversion negatively, associated with relationship conflicts. Founders' openness to experience and agreeableness were positively associated with ventures' performance.

A recent review of the literature on the effects of top executive personality offers a summary of extant findings for each of the Big Five (Holmes Jr et al. 2021). Namely, leaders' openness to experience has been linked with firms' flexibility and performance; conscientiousness has been shown to predict better strategy implementation, but less flexibility and risk-taking; top leaders' extraversion allows them to build social relationships and influence others; agreeableness is associated with greater cohesion and flexibility among subordinates; and neuroticism is associated with lower top management team cohesion, greater relational conflict, and lower firm performance (Peterson et al. 2003).

As noted above, effects of these traits may be less straightforward in the context of online communities because the effects of founders' actions are less direct than those of organizational founders. Founders cannot reward or punish community members, and do not have an official or formal role within the community. Nevertheless, given that community founders are those who establish the community and are the first to engage in it, they have an important role in establishing the community's norms of conduct, such as the amount and pace of communication in the community, the tone of this communication (Garas et al. 2012), responsiveness to others' posts, and the overall atmosphere and climate that come to characterize the community (Cai and Shi 2022; Kaufmann and Vallade 2022). Such norms of conduct typically develop in socialization processes (Schneider et al. 1998) through new members' observations of others' conduct within the community. Given that they are the ones who established the community, founders will have been within the community from its inception to set the tone for others' conduct.

**Online Community Attributes**

Before using online community founders' Big Five traits to predict community attributes and outcomes and attributes, let us first describe some of the key attributes and outcomes that are of interest. First and foremost, much of the literature on online communities focuses on the likelihood that the community will survive over time (e.g., Cunha et al. 2019). Current insights about community sustainability suggest that it is determined, in part, by members' trust in the community and their perceptions of its social usefulness (Lin 2008). Another factor of interest, which also reflects community success, is the amount of activity in the community and its members' engagement in it, including the founder's own retention and activity in the community over time (Cunha et al. 2019; Ellison, Lampe, and Steinfield 2010). Yet other outcomes of interest in online community research involve attributes of the community, as in its size (i.e., number of members), and network structure (e.g., the average number of members with whom members interact). We aim in the present study to complement existing research on these attributes and outcomes by predicting them from founders' personality traits. We draw on the literature on organizational founders and leaders, and on the relationships of founder and leader traits with organizational attributes, to develop predictions about the relationship between online community founders' personality and community attributes. Our research model, developed below, is summarized in Figure 1.

**Figure 1. Research Model**

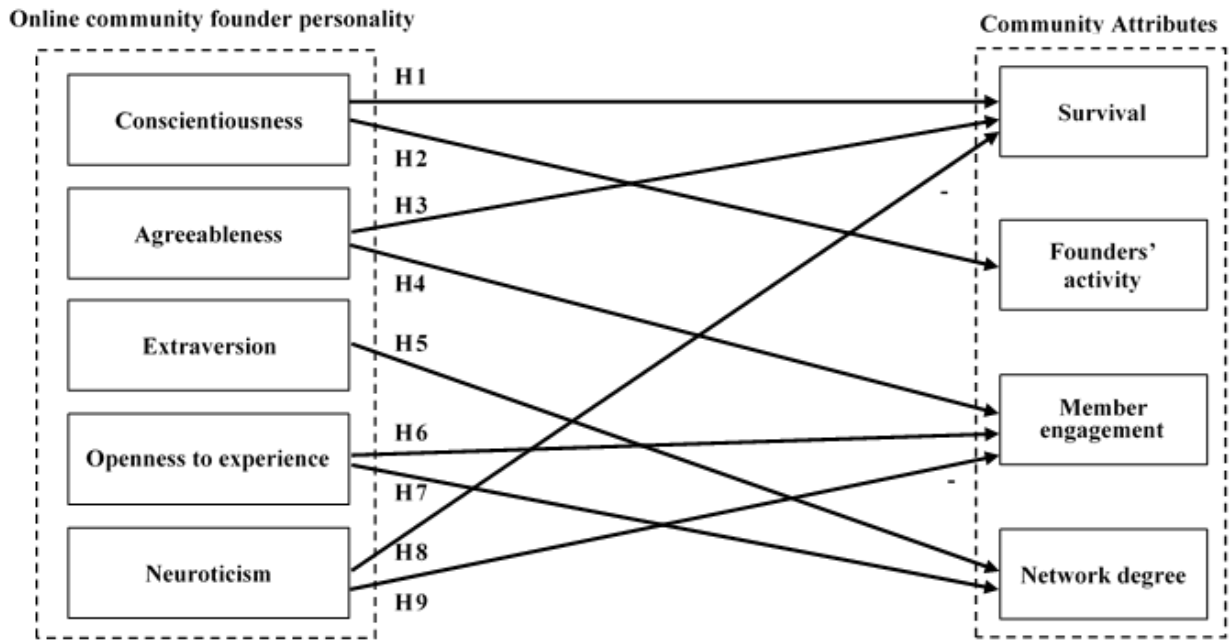

Based on the Big Five trait definitions, there is good reasons to expect links between the traits and online community attributes. First, conscientiousness, which represents individuals' proneness to being dependable, hardworking and methodical has been linked with performance across numerous contexts, and is the most consistent predictor of performance among the Big Five (Barrick and Mount 1991). In organizational research, leader conscientiousness has been linked with effective leadership (Judge et al. 2002), as reflected in followers' in-role and extra-role performance (Hu and Judge 2017). Follower performance is facilitated through conscientious leaders' efforts to ensure followers' role clarity, structure, and goals. In other research the conscientiousness of organizations' top management team was associated with followers' commitment to the organization (Colbert et al. 2014). When applied to the maintenance of an online community, conscientious founders are similarly likely to set clear procedures and guidelines for community members, regularly introduce to the community relevant content, and promptly respond to members' questions, offering reliable and timely information. In doing so, early on in the life of the community, they set a certain standard that other members may follow. Overall, such actions are what allow communities to function and remain active over time. Moreover, of particular relevance to the context of online communities, the dependability and sense of responsibility of conscientious community founders makes it likely that they themselves will remain loyal members of the community. We therefore hypothesize:

> *Hypothesis 1: Community founder conscientiousness will be positively associated with their communities' likelihood to exist one year following its establishment (i.e., community sustainability).*

> *Hypothesis 2: Community founder conscientiousness will be positively associated with the likelihood that founders will remain active members of their communities one year following community establishment.*

Following conscientiousness, leader agreeableness has also been linked with positive follower outcomes (e.g., Derue et al. 2011). Agreeable individuals tend to be friendly and approachable, and agreeable leaders tend to be nurturing toward their followers (Judge and Piccolo 2004). Accordingly, leader agreeableness predicts not only followers' performance, but also their satisfaction with the leader and with their jobs (Derue et al. 2011; Smith and Canger 2004). In the context of online communities, agreeable founders may be more likely to offer favorable and supportive responses to other members of the community, and to contribute, through their communications, to the establishment of a positive and supportive climate within the community. In turn, these should increase the chances that members will choose to remain within the community, and remain actively engaged in it. Thus:

*Hypothesis 3: Community founder agreeableness will be positively associated with their communities' likelihood to exist one year following its establishment (i.e., community sustainability).*

*Hypothesis 4: Community founder agreeableness will be positively associated with members' engagement within the community one year following its establishment.*

Extraverted individuals tend to be outgoing, energetic and communicative. They seek out social stimuli and novel experiences and easily form new relationships with others, and tend to have many friends (Wilson, Harris, and Vazire 2015). Similarly, in research of social networks, individuals' extraversion was associated with their network centrality (Neubert and Taggar 2004). Accordingly, extraverted online community founders could be expected to exhibit socially oriented communications and interact with multiple members of their community. Through their communications in the community, extraverted community founders could be expected to instill a climate that encourages and facilitates interactions among members. As we describe below, this can be tested through the relationship between founders' extraversion and the community's network degree (the average number of individuals with whom community members interact). We therefore hypothesize:

*Hypothesis 5: Community founder extraversion will be positively associated with their communities' network degree.*

Individuals who are open to experiences are creative and open minded, and accordingly value independent thought and innovative ideas (Roccas et al. 2002). They are curious, appreciate novelty, and thus like trying out new things. Similar to extraverted individuals, they seek out stimulation, yet whereas extraverted individuals aim for social stimulation, those who are open to experience focus on intellectual stimulation. When in leadership roles, individuals high on openness to experience highlight novelty in their organizations, which results in their followers' innovative behaviors. Given their preference for new ideas, those who are open to experience should appreciate the diversity of ideas to which they can be exposed by interacting with multiple individuals. Accordingly, open founders can be expected to convey their openness in their communities in their communications and facilitate others' interactions amongst themselves. As such, similar to our prediction about extraversion, founders' openness to experience should be associated with the community's network degree. In addition, given their creativity and originality, founders who are open to experience could be expected to elicit high levels of member engagement in their communities, given the diverse interests they are likely to express in their communications.

> *Hypothesis 6: Community founder openness to experience will be positively associated with members' engagement within the community one year following its establishment.*
>
> *Hypothesis 7: Community founder openness to experience will be positively associated with their communities' network degree.*

Finally, there is also reason to expect relationships between founders' neuroticism and certain attributes of their online community. Neurotic individuals tend to experience negative emotions and stress. Ample research demonstrates its negative effect on individuals' performance in general (e.g., Barrick and Mount 1991; Sosnowska, Hofmans, and De Fruyt 2020) and on leaders' performance in particular (Judge et al. 2002). Those high in neuroticism tend to express their negative emotions, exhibit more sedentary behavior, and exhibit lower levels of activity (Sutin et al. 2016). They are also more likely to behave defensively in their social interactions (Barrett and Pietromonaco 1997). When in leadership roles, neurotic individuals discourage their followers from providing unsolicited input (Walumbwa and Schaubroeck 2009). In the context of online communities, neurotic founders are likely to contribute to an unpleasant atmosphere, through their expression of negative emotions and by modelling impulsive and erratic communication. We can therefore expect founder neuroticism to be associated with their members' activity in the community, as well as their motivation to remain in it over time.

> *Hypothesis 8: Community founder neuroticism will be negatively associated with their communities' likelihood to exist one year following its establishment (i.e., community sustainability).*
>
> *Hypothesis 9: Community founder neuroticism will be negatively associated with members' engagement within the community one year following its establishment.*

## Data and Methods

Our aim is to model the relationship between founders' personality traits and online community attributes. The general idea behind our approach is to use data from a sample of participants to develop an algorithm for estimating community members' personality from their text, and then use the algorithm for estimating a large number of community founders' personalities[1]. To accomplish this, we collected and constructed three data sets: (1) a text-

---

[1] Although a similar approach has been used in a few studies Schwartz, H Andrew, Johannes C Eichstaedt, Margaret L Kern, Lukasz Dziurzynski, Stephanie M Ramones, Megha Agrawal, Achal Shah, Michal Kosinski, David Stillwell, and Martin EP Seligman (2013), "Personality, Gender, and Age in the Language of Social Media: The Open-Vocabulary Approach," *PloS one*, 8 (9), e73791, Yarkoni, Tal (2010), "Personality in 100,000 Words: A Large-Scale Analysis of Personality and Word Use among Bloggers," *Journal of research in personality*, 44 (3), 363-73., we were not able to obtain the data used, which is why we opted to collect our own data.

personality calibration data set, (2) an online communities data set, and (3) online community founders' dataset. We constructed the data sets through the following five steps:

**Step 1: Creating the Calibration Data Set**

We advertised on Reddit that we are looking for users who are willing to fill out a survey about users' personality and text style. One advantage of creating the tool by using Reddit is that it ensures consistency with the platform on which we wish to test our predictions. That is, the context in which we calibrate the tool matches the context in which it is used. We recruited users 18 or older who have published at least 400 words of text on Reddit. Those who agreed to participate, and met our inclusion criteria, were asked to fill out the 20-item mini-IPIP measure of the five-factor model (Donnellan et al. 2006). Data of users who did not complete the entire personality questionnaire, or who provided the same response across all 20 items, were excluded from our analyses (there were only 2 of these). We then extracted each participant's public text on Reddit. Having both users' self-reported personality and their independent texts from Reddit we could construct an algorithm that uses features of users' text to estimate their personality traits. A total of 1,272 users responded to our request, of whom 1,025 provided a valid Reddit username and had at least 400 words of public text, from multiple communities.[2] We extracted features of the text using various language processing methods including LIWC (Tausczik and Pennebaker 2010), affect dictionaries (Warriner, Kuperman, and Brysbaert 2013), and other approaches, as described in Appendix A. In Table 1 we show the means and standard deviations of participants' age, sex, self-reported personality traits, and basic characteristics of the text.

---

[2] We chose 400 words as the minimum to allow sufficient richness in the text, although the results hold when varying this threshold by several hundreds of words.

**Table 1. Properties of the Reddit Calibration Data Set (N=1272)**

|  | Mean | SD |
|---|---|---|
| Sex (0=female, 1=male) |  |  |
| Age | 27.93 | 9.43 |
| Neuroticism | 3.13 | 0.88 |
| Extraversion | 2.70 | 0.94 |
| Openness | 3.99 | 0.74 |
| Agreeableness | 3.82 | 0.79 |
| Conscientiousness | 3.00 | 0.85 |
| Num. of words (per user)* | 143,240 | 370,274 |
| Mean text valence (per user)* | 5.72 | 0.20 |
| Mean text arousal (per user)* | 4.25 | 0.12 |

* Data for these variables were only available for 1025 of the participants.

**Step 2: Using Machine Learning Models to Construct a Text-To-Personality Algorithm**

We next sought to identify features of users' text that can be used for estimating users' personality. We used the Linguistic Inquiry and Word Count software (LIWC, Pennebaker, Francis, and Booth 2001) to calculate several of the features we used, such as the number of negative emotion words, positive emotion words, and social words. We also used emotion norms dictionaries to code affect (Warriner et al. 2013), and bigram and topic analyses (see Appendix A). For robustness, we used the following four well-known statistical models for estimating the personality traits from the text features: (1) General linear model, (2) Random forest model (Biau and Scornet 2016), (3) Gradient boosting machine model (Natekin and Knoll 2013), and, (4) Support-vector machine model (Noble 2006). We used a two-stage feature selection process. First, in line with extant procedures for estimating personality from text (Yarkoni 2010), we selected the 15 features that yielded the highest correlations with each of the five traits. Changing the number of features we used in this stage, up to 100, yielded very similar results (see Appendix A).

In the second feature-selection stage we used a forward-backward stepwise regression to select the optimal set of features of the initial 15. For each of the four statistical models we conducted a grid search to find the optimal parameters for that model (e.g., for the random forest approach, we optimized the number of branches to grow and the number of variables to sample after each split). The goodness-of-fit measures for each model and each personality trait are provided in Table 2. Fit was calculated (1) within the whole sample and, (2) in a process of a K-fold cross-validated sampling and resampling process (Fushiki 2011).

**Table 2. Goodness-of-Fit Measures for the Machine Learning Models**

| Model | Neuroticism | | Extraversion | | Openness | | Agreeableness | | Conscientiousness | |
|---|---|---|---|---|---|---|---|---|---|---|
| | Adj. $R^2$ | Sample-resample Adj. $R^2$ | Adj. $R^2$ | Sample-resample Adj. $R^2$ | Adj. $R^2$ | Sample-resample Adj. $R^2$ | Adj. $R^2$ | Sample-resample Adj. $R^2$ | Adj. $R^2$ | Sample-resample Adj. $R^2$ |
| General linear | 0.25 | 0.18 | 0.23 | 0.14 | 0.24 | 0.12 | 0.24 | 0.15 | 0.22 | 0.15 |
| Support-vector machine | 0.19 | 0.16 | 0.17 | 0.13 | 0.18 | 0.14 | 0.17 | 0.12 | 0.17 | 0.12 |
| Gradient boosting machine | 0.28 | 0.11 | 0.58 | 0.05 | 0.42 | 0.05 | 0.25 | 0.08 | 0.40 | 0.04 |
| Random forest | 0.97 | 0.12 | 0.87 | 0.09 | 0.87 | 0.09 | 0.87 | 0.11 | 0.91 | 0.08 |

Except for the Random Forest model, the variance explained (adjusted R-squared) within the sample is consistent with that observed in other estimations of personality from text (Schwartz et al. 2013), at values ranging from 17% to 58%. The range of variance explained when using sampling-resampling methods, were naturally lower (5% - 18%). This may result from the heterogeneity of the correlations between traits and text features within the sample (we address this issue in the discussion). As additional evidence for the validity of our procedure we also compared the intercorrelations of the Big Five scores of the 1,272 Reddit users who filled out the personality scales with those of the estimated Big Five scores. The correlation matrices of both the surveyed sample and the founders' estimated sample are presented in Appendix C.[3]

**Step 3: Extracting and Processing Online Communities' Data**

To estimate online community founders' personality, and use these estimates to predict community attributes, we extracted community data from Reddit for two time periods: the communities' formation period, and one year later. In line with brand communities (Muñiz & O'Guinn 2001, p. 412), Reddit communities typically consist of consumers that rally around a topic of interest. Indeed, a considerable number of the communities in our sample center around brands (e.g., soccer teams, such as "New Mexico United", the mobile version of the "Call of Duty" video game, and the "Sander Sides" television show). We used online communities first formed between October and November 2018—a total of 8,625 communities (i.e., subreddits on Reddit). We assessed these communities' attributes using users' posts over time, aggregated to the community level. The attributes we assessed include whether the community existed a year following its establishment, its volume of activity, the number of active members, and indicators of its network structure (see Table 3).

---

[3] As can be seen, the patterns of relationships are very similar across the surveyed and estimated scores, with the exception that the correlation between neuroticism and extraversion is negative for the surveyed sample, but positive for the estimated scores. These relationships among the Big Five are also generally consistent with those found in other studies of the five factor model DeYoung, Colin G. (2006), "Higher-Order Factors of the Big Five in a Multi-Informant Sample," *Journal of Personality and Social Psychology*, 91 (6), 1138, Mount, Michael K, Murray R Barrick, Steve M Scullen, and James Rounds (2005), "Higher-Order Dimensions of the Big Five Personality Traits and the Big Six Vocational Interest Types," *Personnel Psychology*, 58 (2), 447-78., except for the correlation between agreeableness and neuroticism, which was positive in our study (for both surveyed and estimated scores), yet is generally negative.

**Table 3. Properties of Communities that Remained Active One-Year Following their Establishment**

|  | N | Mean | SD |
|---|---|---|---|
| Community size | 2,077 | 123.5 | 1342.7 |
| Num. of posts (per community, per month) | 2,077 | 419.8 | 4091.8 |
| % of founders still active | 2,077 | 14.7% | 21.8% |
| Average network degree | 1,376 | 2.13 | 1.27 |
| Network diameter | 2,077 | 2.34 | 3.58 |
| Number of network clusters | 2,077 | 2.89 | 15.64 |

**Step 4: Extracting and Processing Founders' Data**

We first identified founders for each of the 8,625 communities. Founders were the first active members during the community's first two months, up to a maximum of 10.[4] We then extracted for each founder two bodies of text. For estimating founders' personality we used the Reddit texts that existed *prior* to the formation of the community at hand[5], up to the first 2,000 posts. This allowed us to estimate founders' personality independently of their activity within the current community. We estimated founders' personality using the text-to-personality algorithm we developed in Step 2, above. Founders' personality traits were then aggregated to the community level. In other words, we calculated for each community the average of its founders' traits (separately for each of the Big Five). This is consistent with the approach used for studying team personality (e.g., Bradley et al. 2013).

Descriptive statistics of the estimated personality traits, using each of the four estimation models, and key attributes of the sample are presented in Table 4.

---

[4] Note that a user can be a founder of more than a single community.
[5] All of the founders were active on Reddit prior to establishing the current community.

**Table 4. Community Founders' Descriptive Statistics**

|  | General linear | Random forest | Support-vector machine | Gradient boosting machine |
|---|---|---|---|---|
| Neuroticism | 2.99 (0.37) | 2.99 (0.22) | 3.09 (0.12) | 3.02 (0.29) |
| Extraversion | 2.60 (0.36) | 2.62 (0.20) | 2.61 (0.05) | 2.70 (0.01) |
| Openness | 3.97 (0.26) | 3.97 (0.13) | 4.08 (0.01) | 4.03 (0.01) |
| Agreeableness | 3.68 (0.22) | 3.68 (0.18) | 3.86 (0.08) | 3.65 (0.26) |
| Conscientiousness | 2.93 (0.29) | 2.92 (0.17) | 2.92 (0.06) | 2.93 (0.30) |
| Number of founders | | 35,164[*] | | |
| Average number of posts per founder | | 844.4 (777.6) | | |
| Total number of words per founder | | 21,920 (34,938) | | |

[*] Several of the communities had fewer than 10 founders.

**Step 5: Using Founders' Estimated Personality Traits to Predict Community Attributes**

In the final step we ran statistical models using founders' estimated personality traits to predict their communities' attributes. We used the following models:

$$log\left(\frac{\pi_i}{1-\pi_i}\right) = c_0 + \sum_{j=1}^{5} c_j \, PersTrait_j + c_6 n_i + \epsilon_i \quad (1)$$

$$Y_i = b_0 + \sum_{j=1}^{5} b_j \, PersTrait_j + b_6 n_i + \delta_i \quad (2)$$

Equation (1) represents a logistic regression model in which $\pi_i$ denotes the probability of an online community $i$ to remain active one year following its inception. Equation (2) represents a regression model in which $Y_i$ is a property of online community $i$, to be predicted a year following the community's inception. The first community property we considered is the community's sustainability, by testing whether it has remained active one year following the community's inception, i.e., whether there was any activity of any of the community members at the end of the first year. The second property was more specifically focused on whether the community's founders remained active a year after the community's inception. This ratio ranges between 0 (i.e., no founders remained active a year after the community was formed), and 1 (i.e., all founders remained active).

A third and fourth properties were the community's size—defined as the number of active community members for the duration of one month, one year following the community's inception—and the community's engagement, defined as the average activity per person within the community, during that month. To calculate this average, we summed the number of posts and comments produced by all members of the community, during that month, and divided it by the number of active members during that month.

Finally, we assessed a number of attributes of the community's social network of interactions. One index in which we were interested had to do with the degree to which community members interact with each other. We therefore calculated the community's average network degree, which we operationalized as the average of the number of peers with whom each member interacted during the month following the community's inception. We then assessed the network diameter and network degree of centralization. Network diameter reflects the degree to which the community is tightly knit, as indicated in the shortest path between the two most distant people in the network. Network degree centralization refers to the level of inequality in the degree to which community members interact with others. Highly centralized networks are those in which few members interact with many members, whereas the majority of members interact with only few members.

The right-hand side of both equations above has the same structure. The $PersTrait_j$ term represents the Big Five traits averaged across community founders (i.e., founders' average neuroticism, average extraversion, etc.). The five community-averaged founder traits are represented by $PersTrait_j$ (where $j = 1, 2, \ldots, 5$). Finally, because the number of founders per each community varies (see Table 2), we control for the number of founders in the analyses, denoted by $n_i$ in both (1) and (2).

## Results

To assess the robustness of our findings, we use all four statistical models (see Table 4) for our analyses. We lean toward the conservative side and interpret a trait's effect as meaningful for predicting a community attribute only if it is statistically significant across at least three of the four models. This is equivalent to the commonly used ensemble majority vote approach when using several (an ensemble) estimation models (Javid, Alsaedi, and Ghazali 2020). For simplicity of structure, we present our tests of the hypotheses for each type of dependent variable separately. The order in which we test the hypotheses is thus different from the order in which we presented them.

**Founders' Personality Traits and Community Sustainability**

Results of the logistic regression for predicting community sustainability (Equation 1) are provided in Table 5. As can be seen, in line with Hypotheses 1 and 3, both conscientiousness (H1) and agreeableness (H3) are positively related to the community's likelihood of remaining active a year following its inception. A marginal analysis of the logistic model's results shows that a one-point increase of founders' conscientiousness or agreeableness, yields a 5.2% increase in the probability that the community will remain active for at least one year, and a one-point increase in extraversion yields a 1.5% decrease in this probability. As can be seen in Table 5, Hypothesis 8, about the negative relationship between neuroticism and sustainability was not supported.

**Table 5: Predicting Online Communities' Sustainability (N=8,625)**

| Variable | General linear Coeff. (s.e.) | Random forest Coeff. (s.e.) | Support-vector machine Coeff. (s.e.) | Gradient boosting machine Coeff. (s.e.) |
|---|---|---|---|---|
| Intercept | -2.85*** (0.06) | -2.86*** (0.06) | -2.88*** (0.06) | -2.88*** (0.06) |
| Neuroticism | -0.05 (0.03) | 0.01 (0.04) | -0.06 (0.05) | -0.002 (0.04) |
| Extraversion | -0.09*** (0.03) | -0.12*** (0.04) | -0.33*** (0.05) | -0.12*** (0.03) |
| Openness | 0.02 (0.03) | -0.01 (0.03) | 0.06* (0.03) | 0.05 (0.03) |
| Agreeableness | 0.14*** (0.03) | 0.20*** (0.04) | 0.26*** (0.05) | 0.21*** (0.03) |
| Conscientiousness | 0.19*** (0.04) | 0.19*** (0.04) | 0.19*** (0.04) | 0.14*** (0.04) |
| Num. of founders | 0.30*** (0.01) | 0.30*** (0.01) | 0.30*** (0.01) | 0.30*** (0.01) |
| Tjur's $R^2$ (coefficient of determination) | 0.18 | 0.18 | 0.18 | 0.18 |

\* p<0.1, \*\* p<0.05, \*\*\* p<0.01

Beyond the relationships we hypothesized, we also found a significant negative effect of extraversion. This effect may have something to do with the novelty seeking and impulsive tendencies of extraverted individuals (Zuckerman and Glicksohn 2016). Their preference for novel experiences may drive extraverted founders to lose interest in the communities they founded, and turn their attention to new endeavors and adventures, thus neglecting the maintenance of the communities they found. Such a rationale is consistent with the notion that although extraverted individuals are highly social, they are also known to lack high investment and are more likely to lose interest in their current activity, relative to less extraverted individuals (e.g., Flynn, Collins, and Zlatev 2023; Koelega 1992).

Some evidence for this explanation comes from the negative relationship we found between founders' extraversion and the likelihood that they remained in their communities (see below). Beyond the effect of founder personality traits, communities' one-year survival was also positively associated with the number of founders. This may have to do with the possibility that more founders in the community in its early stages may yield higher levels of activity, additional content, and greater responsiveness to other members of the community. All of these could be expected to contribute to communities' survival.

**Founders' Personality Traits and their Own Retention within their Communities**

We use Equation (2) to estimate whether the community-average personality traits predict retention among founders. The results of the estimation are provided in Table 6. As can be seen, in line with Hypothesis 2, founder conscientious predicts the probability that they will remain active within the community. It is also interesting that the bigger the number of founders of the community, the higher the churn rate of founders later after a year of the community's existence.

**Table 6: Founders' Personality Traits and the Ratio of Founders who Remain Active One Year Following the Community's Inception (N=2,077)**

| Variable | General linear Coeff. (s.e.) | Random forest Coeff. (s.e.) | Support-vector machine Coeff. (s.e.) | Gradient boosting machine Coeff. (s.e.) |
|---|---|---|---|---|
| Intercept | 0.34*** (0.01) | 0.33*** (0.01) | 0.33*** (0.01) | 0.33*** (0.01) |
| Neuroticism | -0.01** (0.007) | -0.02 (0.006) | 0.02* (0.01) | -0.01* (0.006) |
| Extraversion | -0.01 (0.008) | 0.001 (0.007) | -0.02** (0.008) | -0.01** (0.006) |
| Openness | 0.006 (0.005) | -0.001 (0.005) | 0.01 (0.006) | 0.02*** (0.005) |
| Agreeableness | 0.01* (0.008) | 0.008 (0.007) | 0.01 (0.01) | 0.02*** (0.006) |
| Conscientiousness | 0.03*** (0.009) | 0.02*** (0.006) | 0.03*** (0.008) | 0.02*** (0.007) |
| Num. of founders | -0.03*** (0.001) | -0.03*** (0.001) | -0.03*** (0.001) | -0.03*** (0.001) |
| Adj. $R^2$ | 0.18 | 0.18 | 0.19 | 0.19 |

* $p<0.1$, ** $p<0.05$, *** $p<0.01$

*Founders Personality Traits and Community Size*

Although we made no predictions about these effects, it is nevertheless interesting to consider the extent to which founders' traits predict community size (Equation 2; Table 7). The analysis for testing these relationships includes only the 2,077 communities that remained active a year following their inception, i.e., our findings are conditioned on the sustainability of the community. We found inconsistent (i.e., in only two of the four models) and weak positive effects of neuroticism and extraversion. More consistently, community size was positively associated with the number of founders.

**Table 7: Founders' Personality Traits and Community Members' Engagement One Year Following Community Inception (N=2,077)**

| Variable | General linear Coeff. (s.e.) | Random forest Coeff. (s.e.) | Support-vector machine Coeff. (s.e.) | Gradient boosting machine Coeff. (s.e.) |
|---|---|---|---|---|
| Intercept | 1.16*** (0.03) | 1.16*** (0.03) | 1.15*** (0.03) | 1.16*** (0.03) |
| Neuroticism | -0.07*** (0.02) | -0.03* (0.02) | -0.10*** (0.03) | -0.001 (0.02) |
| Extraversion | -0.03 (0.02) | 0.01 (0.02) | -0.06** (0.03) | -0.02 (0.02) |
| Openness | 0.02 (0.02) | -0.03* (0.02) | 0.01 (0.01) | 0.05*** (0.02) |
| Agreeableness | 0.08*** (0.03) | 0.06** (0.02) | 0.10*** (0.04) | 0.06*** (0.02) |
| Conscientiousness | 0.04 (0.03) | 0.06*** (0.02) | 0.04 (0.03) | 0.06*** (0.02) |
| Num. of founders | -0.003 (0.004) | -0.003 (0.004) | -0.001 (0.004) | -0.003 (0.004) |
| Adj. $R^2$ | 0.015 | 0.02 | 0.02 | 0.02 |

\* $p<0.1$, \*\* $p<0.05$, \*\*\* $p<0.01$

**Founders' Personality Traits and Community Engagement**

Results of the regression estimation (Equation 2) are provided in Table 8. Here, too, we only included the 2,077 communities that remained active a year after community inception. As can be seen, in line with Hypothesis 4, agreeableness is positively with community member engagement, and in line with Hypothesis 9, neuroticism is negatively associated with it. Hypothesis 6, concerning the effect of founders' openness to experience on members' engagement was not supported.

**Table 8: Founders' Personality Traits and Social Network Degree (N=2,077)**

| Variable | General linear Coeff. (s.e.) | Random forest Coeff. (s.e.) | Support-vector machine Coeff. (s.e.) | Gradient boosting machine Coeff. (s.e.) |
|---|---|---|---|---|
| Intercept | 0.76*** (0.03) | 0.76*** (0.03) | 0.77*** (0.03) | .77*** (0.03) |
| Neuroticism | 0.01 (0.02) | 0.02* (0.01) | -0.012 (0.03) | -0.03* (0.01) |
| Extraversion | 0.07*** (0.02) | 0.08*** (0.02) | 0.10*** (0.02) | 0.07*** (0.01) |
| Openness | -0.01 (0.02) | 0.004 (0.01) | -0.02 (0.02) | -0.01 (0.01) |
| Agreeableness | -0.04** (0.02) | -0.08*** (0.02) | -0.02 (0.03) | -0.04*** (0.01) |
| Conscientiousness | -0.007 (0.02) | 0.01 (0.02) | -0.04* (0.02) | -0.01 (0.02) |
| Num. of founders | 0.04*** (0.003) | 0.04*** (0.003) | 0.04*** (0.003) | 0.04*** (0.003) |
| Adj. $R^2$ | 0.10 | 0.10 | 0.11 | 0.11 |

* $p<0.1$, ** $p<0.05$, *** $p<0.01$

**Founders' Personality Traits and Community Social Network Structure**

Using a model similar to that in Equation 2, and a series of social network structure indicators, we found that founders' personality significantly predicted only social network degree, i.e., the average number of people in the community with whom a member interacts.[6] Results of the analyses for other social network indicators are provided in Appendix A. As can be seen in Table 9, in line with Hypothesis 5, founders' extraversion is positively associated with network degree. The hypothesized effect of founder openness to experience (Hypothesis 7) was not supported. In addition, we found a negative effect for agreeableness. Although we did not hypothesize this effect, it is consistent with the notion that agreeable people invest more in the quality of the relationships they form, rather than in their quantity (Wilson et al. 2015).

---

[6] The distribution of this variable was substantially right skewed, as is commonly documented in the networks literature Barabási, Albert-László and Eric Bonabeau (2003), "Scale-Free Networks," *Scientific american*, 288 (5), 60-69., so we calculated its log which reduced the skewness and helped the distribution approach that of a normal distribution.

## Discussion

Our goal in this study was to link the personality attributes of online community founders to key attributes of their communities. For this purpose, we developed an algorithm for assessing individuals' personality traits from their texts, assessed the traits of 35,164 online community founders, and tested relationships between these traits and a variety of key community attributes. In line with our predictions, we found evidence for meaningful effects of founders' personality on community sustainability, founders' and members' engagement with the community, and the community's social network degree. All five traits help explain certain attributes of the online community. Conscientiousness, agreeableness, and neuroticism (negatively) are associated with community sustainability; Agreeableness, and neuroticism (negatively) were also associated with members' engagement in the community, and conscientiousness and neuroticism (negatively) also explain founders' own engagement with the community. Finally, extraversion and openness to experience (indirectly) were associated with communities' network degree.

**Implications for Research**

Our findings have several implications for research in this and adjacent fields. First, we present a procedure for assessing online users' personality from their texts. The sample with which we applied this procedure is relatively large, thus enhancing the robustness of the resulting tool. Moreover, we make our algorithm and an anonymized version of the data available to other researchers who may be interested in using them for assessing online users' attributes.

Our findings offer an important example of how text analyses of publicly available, user-generated, data could be used for studying individuals' behavior at a large scale (for a review of a similar body of research see (Berger et al. 2020). By applying this approach to the study of community founders we integrate the literature on digital platforms with the literature on top leaders (Hambrick, 2007) in general and on leader personality (e.g., Oreg & Berson, 2018) in particular, to open up new avenues for studying online community leadership and its effects. Moreover, by focusing on community leaders' personality, we extend previous research on community attributes and provide insight into the mechanisms that explain these attributes. Rather than merely describing attributes such as communities' longevity, size, and amount of activity, we explain them through founders' traits.

In addition, our findings extend the literature on leader personality traits by demonstrating their effects in a new context—online communities in digital platforms—and linking them to a new set of outcomes. Whereas some of the effects we found correspond with those established in other contexts, such as the positive correlation between leader conscientiousness and communities' performance (e.g., sustainability and activity within the community), others, such as the negative relationship between extraversion and community sustainability, are less straightforward and may be relevant particularly to the online community context. Our findings are particularly important given the increasing prevalence and popularity of digital platforms, and their impact on day-to-day life. Creating a brand community that could be sustainable and active is a challenge for marketers and brand managers—our findings suggest that user-generated content can be used to make important predictions about community viability, based on an analysis of their founders.

Our findings also highlight the lingering and often unintentional impact that founders have on their communities. Given that people do not choose their personalities, and that personality is relatively stable over time and across situations, our findings raise questions about founders' ability to change their communities by design. This is not to say that founders have no control in shaping their communities, given that personality is only one of many factors that contribute to community attributes. Yet what we find does suggest that founders' control may nevertheless be restricted by their personal dispositions.

The research approach we used in this study provides a useful means of overcoming the difficulty in obtaining self-reports of founders' and top leaders' personality. Given this difficulty, the number of studies that use top leaders' ratings of their own personality is relatively limited. The approach we used allows us to assess the personalities of an almost unlimited number of leaders, not only for the study of online communities, but any forum of organizations in which leaders' texts are available.

**Implications for Practice**

Using the approach we presented here, digital platforms, marketers and brand managers can assess founders' personality and assess its fit to the community's goals. They, and the founders themselves, can similarly assess the degree of fit among founders. Although founders will not be able to alter their personality, awareness to its role can be used to compensate for it by recruiting additional founders with complementary traits, and by deliberately taking action (e.g., encourage activity in the community) that may not be readily consistent with their own traits.

For communities soon after inception, platforms, marketers and brand managers can also use our approach to predict communities' sustainability and other attributes. This will allow them to introduce policies and incentives that could compensate for expected outcomes, which may be inconsistent with their preferences for the community. For example, if a given community's founders' exhibit low levels of conscientiousness, such that the expected sustainability is also low, platforms or brand managers can consider offering incentives to encourage founders to invest more effort and time in managing the community, or recruit other founders who show better fit with the community's intended goals. This paper also contributes to the literature managers' use of artificial intelligence in their daily decision making (Berente et al. 2021). We show that AI tools can be useful for predicting the behavior of individuals and groups in digital platforms, and can thus help managers evaluate and predict the performance of the digital platforms they rely on, and make decisions accordingly. In summary, our findings should encourage platforms and firms to acknowledge the role of founders in general, and founders' personality traits in particular, in shaping the future of online communities, and thus design improved policies for optimizing online communities' attributes and sustainability.

**Limitations and Directions for Future Research**

One limitation of our study pertains to the limited role of personality in predicting outcomes, given the relatively modest effect sizes. That said, the effects we obtained are comparable with those in other personality studies, and are nevertheless valuable given the complexity of the outcomes we are predicting. Second, given the correlational design of our study, we can only assume the causal nature of the relationships we obtained. Although our findings are consistent with theory and previous research about the role of founder and leader personality in shaping organizations, and the timing in which we measured founders' personality and the outcomes is consistent with our arguments that founder personality is what drives the community's attributes, the correlations we obtained may nevertheless result from other causal pathways. For example, founder personality is likely associated with the topic to which an online community is dedicated, and, in turn, certain topics may lend themselves more or less readily to certain types of attributes. In practice, it is likely that more than a single causal pathway contributes to the relationship between founder personality and community attributes. Furthermore, whereas we demonstrate relationships between founders' personality and communities' attributes, there would be value in uncovering the process through which these relationships occur. We know from some research of more traditional organizations, for example, that leader behaviors can mediate the effects of leader traits on organizational attributes (e.g., Oreg & Berson, 2018). Future research could look into such possibilities specifically in the context of online communities.

Another limitation involves the external validity of our findings. Although our theoretical arguments should hold for other contexts and platforms, our findings were obtained in the particular context of Reddit and were not restricted to brand communities. Nevertheless, Reddit communities are typically very similar in structure and function as typical brand communities ((Muniz Jr and O'guinn 2001). Future research could test our predictions in other platforms that host online communities, including those restricted to brand communities, and consider other types of members and leaders, such as community moderators, community experts, and members that are categorized based on the level of contribution to the community.

Through the methodology we used, other research can look into other sets of founder attributes. Whereas we focused on personality traits, others can consider focusing on the role of founders' personal values (e.g., Schwartz, 1992), which involve a different aspect of personality, and has also been shown to predict important personal and organizational outcomes.

## Appendix A. Extraction of Features from the Text

Each founder's posts and comments were collated into a single body of text which was then processed by text analysis tools. The processing was done using three tools: (1) The Linguistic Inquiry and Word Count (LIWC), (2) Emotion word dictionaries, and (3) bigram analyses. For the LIWC approach we used the LIWC 2015 software that produces variables using pre-coded categories (Pennebaker et al. 2015). LIWC is a collection of dictionaries on a variety of topics. For example, the category "affect" measures the percent of words in a given body of text that are in the affect-related category (e.g., "happy", "cried"). For a complete list of variables see (Pennebaker et al. 2015).

To code emotion for each user's text, we used the emotion dictionaries from Warriner (2013). Each word in each user's text was matched its level of valence and arousal as indicated in the dictionaries. Then, for each text, the mean and standard deviation of the valence of all words in the text were calculated and added as features of the text. Finally, we collected bigrams (i.e., word couples) across all users' text. For each user we collected the 50 most frequently used bigrams. Overall, across all users, 24,109 bigrams were collected. We then kept only the bigrams that appeared in more than 100 users. The end result was that we used 602 bigrams as text features (e.g., "rich people", and "start playing"). The collection of LIWC, emotion dictionary, and bigram features were used as predictors in our study's model for estimating personality from the text. We also note that we considered other types of text features for estimating personality, such as Latent Dirichlet Allocation algorithms for extracting topics from the text. These other features did not yield a meaningful improvement over the main features we used.

## Appendix B. Testing the role of personality for predicting other social networks indicators

We present here the results of our estimations for network diameter, degree centralization and closeness centralization. As can be seen in Tables B.1 to B.3, we do not observe any meaningful effects of personality traits on these indicators.

**Table B.1: Founders' Personality Traits and Social Network Diameter (N=2,077)**

| Variables | GLM β | GLM p | RF β | RF p | SVM β | SVM p | GBM β | GBM p |
|---|---|---|---|---|---|---|---|---|
| Intercept | -.26 | .158 | -.28 | .135 | -.23 | .202 | -.28 | .133 |
| Neuroticism | .26 | .021 | .16 | .115 | .19 | .279 | .07 | .510 |
| Agreeableness | -.21 | .134 | -.21 | .087 | -.13 | .524 | -.19 | .050 |
| Opennness | .03 | .729 | .04 | .704 | .06 | .571 | .06 | .470 |
| Extraversion | .15 | .265 | -.02 | .845 | .16 | .245 | .09 | .347 |
| Conscientiousness | -.03 | .864 | -.04 | .730 | -.15 | .306 | -.09 | .410 |
| # of founders | .36 | 0 | .36 | 0 | .35 | 0 | .36 | 0 |
| N | 2077 | | 2077 | | 2077 | | 2077 | |
| $R^2$ | 0.11 | | 0.11 | | 0.11 | | 0.11 | |

**Table B.2: Founders' Personality Traits and Social Network Degree Centralization (N=2,077)**

| Variables | GLM β | GLM p | RF β | RF p | SVM β | SVM p | GBM β | GBM p |
|---|---|---|---|---|---|---|---|---|
| Intercept | 0.14 | 0.000 | 0.14 | 0.000 | 0.14 | 0.000 | 0.14 | 0.000 |
| Neuroticism | -0.00 | 0.727 | 0.002 | 0.790 | -0.003 | 0.807 | -0.01 | 0.360 |
| Agreeableness | 0.02 | 0.136 | -0.01 | 0.454 | 0.01 | 0.653 | 0.0002 | 0.980 |
| Openness | 0.002 | 0.761 | 0.01 | 0.380 | 0.001 | 0.907 | 0.001 | 0.932 |
| Extraversion | 0.003 | 0.723 | 0.02 | 0.010 | 0.01 | 0.577 | 0.02 | 0.0002 |
| Conscientiousness | 0.01 | 0.341 | 0.01 | 0.149 | 0.01 | 0.262 | 0.01 | 0.518 |
| # of founders | 0.004 | 0.008 | 0.004 | 0.008 | 0.004 | 0.007 | 0.004 | 0.014 |
| N | 1315 | | 1315 | | 1315 | | 1315 | |
| $R^2$ | 0.01 | | 0.02 | | 0.01 | | 0.02 | |

**Table B.3: Founders' Personality Traits and Social Network Closeness Centralization (N=2,077)**

|  | GLM | | RF | | SVM | | GBM | |
|---|---|---|---|---|---|---|---|---|
| variables | β | p | β | p | β | p | β | p |
| Intercept | 0.58 | 0 | 0.58 | 0 | 0.59 | 0 | 0.59 | 0 |
| Neuroticism | -0.006 | 0.785 | -0.01 | 0.475 | -0.05 | 0.140 | 0.001 | 0.938 |
| Agreeableness | -0.05 | 0.080 | 0.02 | 0.442 | 0.07 | 0.060 | 0.01 | 0.691 |
| Opennness | 0.01 | 0.577 | 0.02 | 0.221 | 0.01 | 0.732 | 0.02 | 0.149 |
| Extraversion | -0.03 | 0.218 | 0.01 | 0.571 | -0.02 | 0.433 | 0.03 | 0.045 |
| Conscientiousness | 0.04 | 0.233 | 0.03 | 0.129 | 0.01 | 0.724 | 0.01 | 0.503 |
| # of founders | -0.03 | 0 | -0.03 | 0 | -0.03 | 0 | -0.03 | 0 |
| N | 1042 | | 1042 | | 1042 | | 1042 | |
| $R^2$ | 0.04 | | 0.05 | | 0.04 | | 0.05 | |

# Appendix C. Big Five Intercorrelation for surveyed and founders' estimated sample

The following are intercorrelations among the Big Five for the surveyed sample (Table C.1) and the sample of founders, for which we estimated personality scores using the four estimation methods (Table C.2).

## Table C.1 Big Five Intercorrelations of the Reddit Calibration (Surveyed) Data Set

|  | Neuroticism | Extraversion | Conscientiousness | Openness |
|---|---|---|---|---|
| Extraversion | -.11*** |  |  |  |
| Conscientiousness | -.22*** | -.01 |  |  |
| Openness | .01 | .20*** | -.04 |  |
| Agreeableness | .12*** | .29*** | .00 | .19*** |

*** p<0.01
Note: Data for these variables were only available for 1,272 of the participants who completed the survey.

## Table C.2 Intercorrelations of Founders' Estimated Big Five

|  | Estimation | Neuroticism | Extraversion | Conscientiousness | Openness |
|---|---|---|---|---|---|
| Extraversion | GLM | .01 |  |  |  |
|  | RF | .15*** |  |  |  |
|  | SVM | .31*** |  |  |  |
|  | GBM | .15*** |  |  |  |
| Conscientiousness | GLM | -.17*** | .11*** |  |  |
|  | RF | -.22*** | .25*** |  |  |
|  | SVM | -.30*** | .27*** |  |  |
|  | GBM | -.27*** | .23*** |  |  |
| Openness | GLM | -.05*** | .07*** | .03*** |  |
|  | RF | -.07*** | .25*** | .18*** |  |
|  | SVM | -.19*** | .23*** | .25*** |  |
|  | GBM | -.18*** | .08*** | .06*** |  |
| Agreeableness | GLM | .40*** | .21*** | .18*** | .04*** |
|  | RF | .37*** | .53*** | .28*** | .17*** |
|  | SVM | .64*** | .67*** | .22*** | .08*** |
|  | GBM | .23*** | .36*** | .30*** | -.02*** |

*** p<0.01
Note: Data for these variables were only available for 35,164 of the participants.

**References**

Adamopoulos, Panagiotis, Anindya Ghose, and Vilma Todri (2018), "The Impact of User Personality Traits on Word of Mouth: Text-Mining Social Media Platforms," *Information systems research*, 29 (3), 612-40.

Algesheimer, René, Utpal M Dholakia, and Andreas Herrmann (2005), "The Social Influence of Brand Community: Evidence from European Car Clubs," *Journal of marketing*, 69 (3), 19-34.

Antonacci, Grazia, Andrea Fronzetti Colladon, Alessandro Stefanini, and Peter Gloor (2017), "It Is Rotating Leaders Who Build the Swarm: Social Network Determinants of Growth for Healthcare Virtual Communities of Practice," *Journal of Knowledge Management*.

Arazy, Ofer, Johannes Daxenberger, Hila Lifshitz-Assaf, Oded Nov, and Iryna Gurevych (2016), "Turbulent Stability of Emergent Roles: The Dualistic Nature of Self-Organizing Knowledge Coproduction," *Information systems research*, 27 (4), 792-812.

Arazy, Ofer, Oded Nov, Raymond Patterson, and Lisa Yeo (2011), "Information Quality in Wikipedia: The Effects of Group Composition and Task Conflict," *Journal of management information systems*, 27 (4), 71-98.

Arazy, Ofer, Felipe Ortega, Oded Nov, Lisa Yeo, and Adam Balila (2015), "Functional Roles and Career Paths in Wikipedia," in *Proceedings of the 18th ACM conference on computer supported cooperative work & social computing*, 1092-105.

Bagozzi, Richard P and Utpal M Dholakia (2006), "Antecedents and Purchase Consequences of Customer Participation in Small Group Brand Communities," *International Journal of research in Marketing*, 23 (1), 45-61.

Barabási, Albert-László and Eric Bonabeau (2003), "Scale-Free Networks," *Scientific american*, 288 (5), 60-69.

Barrett, Lisa Feldman and Paula R Pietromonaco (1997), "Accuracy of the Five-Factor Model in Predicting Perceptions of Daily Social Interactions," *Personality and Social Psychology Bulletin*, 23 (11), 1173-87.

Barrett, Michael, Eivor Oborn, and Wanda Orlikowski (2016), "Creating Value in Online Communities: The Sociomaterial Configuring of Strategy, Platform, and Stakeholder Engagement," *Information systems research*, 27 (4), 704-23.

Barrick, Murray R and Michael K Mount (1991), "The Big Five Personality Dimensions and Job Performance: A Meta-Analysis," *Personnel psychology*, 44 (1), 1-26.

Bateman, Patrick J, Peter H Gray, and Brian S Butler (2011), "Research Note—the Impact of Community Commitment on Participation in Online Communities," *Information systems research*, 22 (4), 841-54.

Berente, Nicholas, Bin Gu, Jan Recker, and Radhika Santhanam (2021), "Managing Artificial Intelligence," *Mis Quarterly*, 45 (3), 1433-50.

Berger, Jonah, Ashlee Humphreys, Stephan Ludwig, Wendy W Moe, Oded Netzer, and David A Schweidel (2020), "Uniting the Tribes: Using Text for Marketing Insight," *Journal of marketing*, 84 (1), 1-25.

Berson, Yair and Shaul Oreg (2016), "The Role of School Principals in Shaping Children's Values," *Psychological Science*, 27 (12), 1539-49.

Berson, Yair, Shaul Oreg, and Taly Dvir (2008), "Ceo Values, Organizational Culture and Firm Outcomes," *Journal of Organizational Behavior: the International Journal of Industrial, Occupational and Organizational Psychology and Behavior*, 29 (5), 615-33.

Biau, Gérard and Erwan Scornet (2016), "A Random Forest Guided Tour," *Test*, 25 (2), 197-227.

Bradley, Bret H, Anthony C Klotz, Bennett E Postlethwaite, and Kenneth G Brown (2013), "Ready to Rumble: How Team Personality Composition and Task Conflict Interact to Improve Performance," *Journal of Applied Psychology*, 98 (2), 385.

Brogi, Stefano (2014), "Online Brand Communities: A Literature Review," *Procedia-Social and Behavioral Sciences*, 109, 385-89.

Butler, Brian S, Patrick J Bateman, Peter H Gray, and E Ilana Diamant (2014), "An Attraction–Selection–Attrition Theory of Online Community Size and Resilience," *Mis Quarterly*, 38 (3), 699-729.

Cai, Yang and Wendian Shi (2022), "The Influence of the Community Climate on Users' Knowledge-Sharing Intention: The Social Cognitive Theory Perspective," *Behaviour & information technology*, 41 (2), 307-23.

Cervone, Daniel and Lawrence A Pervin (2015), *Personality: Theory and Research*, John Wiley & Sons.

Chang, Aihwa, Sara H Hsieh, and Frances Lin (2013), "Personality Traits That Lead Members of Online Brand Communities to Participate in Information Sending and Receiving," *International Journal of Electronic Commerce*, 17 (3), 37-62.

Colbert, Amy E, Murray R Barrick, and Bret H Bradley (2014), "Personality and Leadership Composition in Top Management Teams: Implications for Organizational Effectiveness," *Personnel psychology*, 67 (2), 351-87.

Cunha, Tiago, David Jurgens, Chenhao Tan, and Daniel Romero (2019), "Are All Successful Communities Alike? Characterizing and Predicting the Success of Online Communities," in *The World Wide Web Conference*, 318-28.

de Jong, Ad, Michael Song, and Lisa Z Song (2013), "How Lead Founder Personality Affects New Venture Performance: The Mediating Role of Team Conflict," *Journal of Management*, 39 (7), 1825-54.

DeNeve, Kristina M and Harris Cooper (1998), "The Happy Personality: A Meta-Analysis of 137 Personality Traits and Subjective Well-Being," *Psychological bulletin*, 124 (2), 197.

Derue, D Scott, Jennifer D Nahrgang, Ned ED Wellman, and Stephen E Humphrey (2011), "Trait and Behavioral Theories of Leadership: An Integration and Meta-Analytic Test of Their Relative Validity," *Personnel psychology*, 64 (1), 7-52.

DeYoung, Colin G. (2006), "Higher-Order Factors of the Big Five in a Multi-Informant Sample," *Journal of Personality and Social Psychology*, 91 (6), 1138.

Digman, John M (1990), "Personality Structure: Emergence of the Five-Factor Model," *Annual review of psychology*, 41 (1), 417-40.

Donnellan, M Brent, Frederick L Oswald, Brendan M Baird, and Richard E Lucas (2006), "The Mini-Ipip Scales: Tiny-yet-Effective Measures of the Big Five Factors of Personality," *Psychological assessment*, 18 (2), 192.

Ellison, Nicole B, Cliff Lampe, and Charles Steinfield (2010), "With a Little Help from My Friends: How Social Network Sites Affect Social Capital Processes," *A networked self*, 132-53.

Flynn, Francis J, Hanne Collins, and Julian Zlatev (2023), "Are You Listening to Me? The Negative Link between Extraversion and Perceived Listening," *Personality and Social Psychology Bulletin*, 49 (6), 837-51.

Foote, Jeremy and Noshir Contractor (2018), "The Behavior and Network Position of Peer Production Founders," in *International Conference on Information*, Springer, 99-106.

Foote, Jeremy, Darren Gergle, and Aaron Shaw (2017), "Starting Online Communities: Motivations and Goals of Wiki Founders," in *Proceedings of the 2017 CHI Conference on Human Factors in Computing Systems*, 6376-80.

Fournier, Susan and Lara Lee (2009), "Getting Brand Communities Right," *Harvard business review*, 87 (4), 105-11.

Fushiki, Tadayoshi (2011), "Estimation of Prediction Error by Using K-Fold Cross-Validation," *Statistics and Computing*, 21 (2), 137-46.

Gamache, Daniel L, Gerry McNamara, Michael J Mannor, and Russell E Johnson (2015), "Motivated to Acquire? The Impact of Ceo Regulatory Focus on Firm Acquisitions," *Academy of Management Journal*, 58 (4), 1261-82.

Garas, Antonios, David Garcia, Marcin Skowron, and Frank Schweitzer (2012), "Emotional Persistence in Online Chatting Communities," *Scientific Reports*, 2 (1), 402.

Gazit, Tali (2021), "Exploring Leadership in Facebook Communities: Personality Traits and Activities," in *Proceedings of the 54th Hawaii International Conference on System Sciences*, 3027.

Godes, David and Dina Mayzlin (2004), "Using Online Conversations to Study Word-of-Mouth Communication," *Marketing Science*, 23 (4), 545-60.

Gupta, Abhinav, Forrest Briscoe, and Donald C Hambrick (2018), "Evenhandedness in Resource Allocation: Its Relationship with Ceo Ideology, Organizational Discretion, and Firm Performance," *Academy of Management Journal*, 61 (5), 1848-68.

Hambrick, Donald C (2007), "Upper Echelons Theory: An Update," Vol. 32, Academy of Management Briarcliff Manor, NY 10510, 334-43.

Hambrick, Donald C and Phyllis A Mason (1984), "Upper Echelons: The Organization as a Reflection of Its Top Managers," *Academy of management review*, 9 (2), 193-206.

Hinds, David and Ronald M Lee (2008), "Social Network Structure as a Critical Success Condition for Virtual Communities," in *Proceedings of the 41st Annual Hawaii International Conference on System Sciences (HICSS 2008)*, IEEE, 323-23.

Hirsh, Jacob B, Colin G DeYoung, and Jordan B Peterson (2009), "Metatraits of the Big Five Differentially Predict Engagement and Restraint of Behavior," *Journal of personality*, 77 (4), 1085-102.

Holmes Jr, R Michael, Michael A Hitt, Pamela L Perrewe, Joshua C Palmer, and Gonzalo Molina-Sieiro (2021), "Building Cross-Disciplinary Bridges in Leadership: Integrating Top Executive Personality and Leadership Theory and Research," *The Leadership Quarterly*, 32 (1), 101490.

Hu, Jia and Timothy A Judge (2017), "Leader–Team Complementarity: Exploring the Interactive Effects of Leader Personality Traits and Team Power Distance Values on Team Processes and Performance," *Journal of Applied Psychology*, 102 (6), 935.

Javid, Irfan, Ahmed Khalaf Zager Alsaedi, and Rozaida Ghazali (2020), "Enhanced Accuracy of Heart Disease Prediction Using Machine Learning and Recurrent Neural Networks Ensemble Majority Voting Method," *International Journal of Advanced Computer Science and Applications*, 11 (3).

Judge, Timothy A, Joyce E Bono, Remus Ilies, and Megan W Gerhardt (2002), "Personality and Leadership: A Qualitative and Quantitative Review," *Journal of applied psychology*, 87 (4), 765.

Judge, Timothy A and Ronald F Piccolo (2004), "Transformational and Transactional Leadership: A Meta-Analytic Test of Their Relative Validity," *Journal of applied psychology*, 89 (5), 755.

Kaufmann, Renee and Jessalyn I Vallade (2022), "Exploring Connections in the Online Learning Environment: Student Perceptions of Rapport, Climate, and Loneliness," *Interactive Learning Environments*, 30 (10), 1794-808.

Koelega, Harry S (1992), "Extraversion and Vigilance Performance: 30 Years of Inconsistencies," *Psychological bulletin*, 112 (2), 239.

Kozinets, Robert V, Kristine De Valck, Andrea C Wojnicki, and Sarah JS Wilner (2010), "Networked Narratives: Understanding Word-of-Mouth Marketing in Online Communities," *Journal of marketing*, 74 (2), 71-89.

Kraut, Robert E and Andrew T Fiore (2014), "The Role of Founders in Building Online Groups," in *Proceedings of the 17th ACM conference on Computer supported cooperative work & social computing*, 722-32.

Larson, Lisa M, Patrick J Rottinghaus, and Fred H Borgen (2002), "Meta-Analyses of Big Six Interests and Big Five Personality Factors," *Journal of Vocational Behavior*, 61 (2), 217-39.

Lazar, Jonathan and Jennifer Preece (2002), *Social Considerations in Online Communities: Usability, Sociability, and Success Factors*, na.

Lin, Hsiu-Fen (2008), "Determinants of Successful Virtual Communities: Contributions from System Characteristics and Social Factors," *Information & Management*, 45 (8), 522-27.

Lu, Xianghua, Chee Wei Phang, and Jie Yu (2011), "Encouraging Participation in Virtual Communities through Usability and Sociability Development: An Empirical

Investigation," *ACM SIGMIS Database: The DATABASE for Advances in Information Systems*, 42 (3), 96-114.

Manchanda, Puneet, Grant Packard, and Adithya Pattabhiramaiah (2015), "Social Dollars: The Economic Impact of Customer Participation in a Firm-Sponsored Online Customer Community," *Marketing Science*, 34 (3), 367-87.

McAlexander, James H, John W Schouten, and Harold F Koenig (2002), "Building Brand Community," *Journal of marketing*, 66 (1), 38-54.

Mount, Michael K, Murray R Barrick, Steve M Scullen, and James Rounds (2005), "Higher-Order Dimensions of the Big Five Personality Traits and the Big Six Vocational Interest Types," *Personnel Psychology*, 58 (2), 447-78.

Muniz Jr, Albert M and Thomas C O'guinn (2001), "Brand Community," *Journal of consumer research*, 27 (4), 412-32.

Nadkarni, Sucheta and POL Herrmann (2010), "Ceo Personality, Strategic Flexibility, and Firm Performance: The Case of the Indian Business Process Outsourcing Industry," *Academy of Management Journal*, 53 (5), 1050-73.

Natekin, Alexey and Alois Knoll (2013), "Gradient Boosting Machines, a Tutorial," *Frontiers in neurorobotics*, 7, 21.

Neubert, Mitchell J and Simon Taggar (2004), "Pathways to Informal Leadership: The Moderating Role of Gender on the Relationship of Individual Differences and Team Member Network Centrality to Informal Leadership Emergence," *The Leadership Quarterly*, 15 (2), 175-94.

Noble, William S (2006), "What Is a Support Vector Machine?," *Nature biotechnology*, 24 (12), 1565-67.

Oreg, Shaul and Yair Berson (2018), "The Impact of Top Leaders' Personalities: The Processes through Which Organizations Become Reflections of Their Leaders," *Current Directions in Psychological Science*, 27 (4), 241-48.

Pennebaker, James W, Ryan L Boyd, Kayla Jordan, and Kate Blackburn (2015),The Development and Psychometric Properties of Liwc2015.

Pennebaker, James W, Martha E Francis, and Roger J Booth (2001), "Linguistic Inquiry and Word Count: Liwc 2001," *Mahway: Lawrence Erlbaum Associates*, 71 (2001), 2001.

Peterson, Randall S, D Brent Smith, Paul V Martorana, and Pamela D Owens (2003), "The Impact of Chief Executive Officer Personality on Top Management Team Dynamics: One Mechanism by Which Leadership Affects Organizational Performance," *Journal of applied Psychology*, 88 (5), 795.

Reddit "Reddit.Com," http://www.Reddit.com.

Robert Jr, Lionel P and Daniel M Romero (2017), "The Influence of Diversity and Experience on the Effects of Crowd Size," *Journal of the Association for Information Science and Technology*, 68 (2), 321-32.

Roccas, Sonia, Lilach Sagiv, Shalom H Schwartz, and Ariel Knafo (2002), "The Big Five Personality Factors and Personal Values," *Personality and Social Psychology Bulletin*, 28 (6), 789-801.

Schau, Hope Jensen, Albert M Muñiz Jr, and Eric J Arnould (2009), "How Brand Community Practices Create Value," *Journal of marketing*, 73 (5), 30-51.

Schneider, Benjamin (1987), "The People Make the Place," *Personnel psychology*, 40 (3), 437-53.

Schneider, Benjamin, D Brent Smith, Sylvester Taylor, and John Fleenor (1998), "Personality and Organizations: A Test of the Homogeneity of Personality Hypothesis," *Journal of Applied Psychology*, 83 (3), 462.

Schwartz, H Andrew, Johannes C Eichstaedt, Margaret L Kern, Lukasz Dziurzynski, Stephanie M Ramones, Megha Agrawal, Achal Shah, Michal Kosinski, David Stillwell, and Martin EP Seligman (2013), "Personality, Gender, and Age in the Language of Social Media: The Open-Vocabulary Approach," *PloS one*, 8 (9), e73791.

Smith, Mark Alan and Jonathan M Canger (2004), "Effects of Supervisor "Big Five" Personality on Subordinate Attitudes," *Journal of business and psychology*, 18 (4), 465-81.

Sosnowska, Joanna, Joeri Hofmans, and Filip De Fruyt (2020), "Revisiting the Neuroticism–Performance Link: A Dynamic Approach to Individual Differences," *Journal of Occupational and Organizational Psychology*, 93 (2), 495-504.

Sutin, Angelina R, Yannick Stephan, Martina Luchetti, Ashley Artese, Atsushi Oshio, and Antonio Terracciano (2016), "The Five-Factor Model of Personality and Physical Inactivity: A Meta-Analysis of 16 Samples," *Journal of research in personality*, 63, 22-28.

Tausczik, Yla R and James W Pennebaker (2010), "The Psychological Meaning of Words: Liwc and Computerized Text Analysis Methods," *Journal of language and social psychology*, 29 (1), 24-54.

Walumbwa, Fred O and John Schaubroeck (2009), "Leader Personality Traits and Employee Voice Behavior: Mediating Roles of Ethical Leadership and Work Group Psychological Safety," *Journal of applied psychology*, 94 (5), 1275.

Wang, Gang, R Michael Holmes Jr, In-Sue Oh, and Weichun Zhu (2016), "Do Ceos Matter to Firm Strategic Actions and Firm Performance? A Meta-Analytic Investigation Based on Upper Echelons Theory," *Personnel Psychology*, 69 (4), 775-862.

Warriner, Amy Beth, Victor Kuperman, and Marc Brysbaert (2013), "Norms of Valence, Arousal, and Dominance for 13,915 English Lemmas," *Behavior research methods*, 45 (4), 1191-207.

Wilson, Robert E, Kelci Harris, and Simine Vazire (2015), "Personality and Friendship Satisfaction in Daily Life: Do Everyday Social Interactions Account for Individual Differences in Friendship Satisfaction?," *European Journal of Personality*, 29 (2), 173-86.

Xia Liu, Angela, Yilin Li, and Sean Xin Xu (2021), "Assessing the Unacquainted: Inferred Reviewer Personality and Review Helpfulness," *Mis Quarterly*, 45 (3).

Yarkoni, Tal (2010), "Personality in 100,000 Words: A Large-Scale Analysis of Personality and Word Use among Bloggers," *Journal of research in personality*, 44 (3), 363-73.

Zhu, Dong Hong, Hui Sun, and Ya Ping Chang (2016), "Effect of Social Support on Customer Satisfaction and Citizenship Behavior in Online Brand Communities: The Moderating Role of Support Source," *Journal of Retailing and Consumer Services*, 31, 287-93.

Zuckerman, Marvin and Joseph Glicksohn (2016), "Hans Eysenck's Personality Model and the Constructs of Sensation Seeking and Impulsivity," *Personality and Individual Differences*, 103, 48-52.